\documentclass[%
 aip,
 jmp,%
 amsmath,amssymb,
reprint,%
]{revtex4-1}

\usepackage{graphicx}
\usepackage{epstopdf}
\usepackage{dcolumn}
\usepackage{bm}

\begin{document}


\title{Unsupervised machine learning for detection of phase transitions in off-lattice systems II. Applications}
\author{R. B. Jadrich} 
\author{B. A. Lindquist} 
\affiliation{McKetta Department of Chemical Engineering, University of Texas at Austin, Austin, Texas 78712, USA}
\author{W. D. Pi\~neros}
\affiliation{Department of Chemistry, University of Texas at Austin, Austin, Texas 78712, USA}
\author{D. Banerjee}
\affiliation{McKetta Department of Chemical Engineering, University of Texas at Austin, Austin, Texas 78712, USA}
\author{T. M. Truskett} 
 \email{truskett@che.utexas.edu}
\affiliation{McKetta Department of Chemical Engineering, University of Texas at Austin, Austin, Texas 78712, USA}
\affiliation{Department of Physics, University of Texas at Austin, Austin, Texas 78712, USA}

\date{\today}

\begin{abstract}
We outline how principal component analysis (PCA) can be applied to particle configuration data to detect a variety of phase transitions in off-lattice systems, both in and out of equilibrium. Specifically, we discuss its application to study 1) the nonequilibrium random organization (RandOrg) model that exhibits a phase transition from quiescent to steady-state behavior as a function of density, 2) orientationally and positionally driven equilibrium phase transitions for hard ellipses, and 3) compositionally driven demixing transitions in the non-additive binary Widom-Rowlinson mixture.
\end{abstract}

\maketitle

\section{Introduction}

Principal component analysis (PCA) is a simple and widely used unsupervised machine learning tool for dimensionality reduction.~\cite{br_and_ml_book,stat_learning_book,pca_guide} Perhaps the most common application of PCA is for the lossy compression of images. One popular demonstration is the analysis of facial images, leading to the aptly named ``eigenfaces'' that capture collective attributes of facial structure.~\cite{br_and_ml_book,stat_learning_book} Only a subset of the eigenfaces--much fewer than the na\"{i}ve dimensionality of the problem--are required to recover the salient aspects of facial images by simple linear combination. Another routine use is in natural language processing, where PCA is employed to shrink the data dimensionality down from the large number of words appearing in a data set or in a dictionary.~\cite{br_and_ml_book,stat_learning_book} Use of the resultant lower dimensional representation greatly improves the development of predictive models to classify text documents. 

The combined power and simplicity of PCA has made it a popular tool in the biological and physical sciences as well. For example, DNA microarray data is routinely treated with PCA to reduce the high dimensionality of the problem in order to identify unique gene expression states across various experimental conditions.~\cite{stat_learning_book, dna_microarray_chapter} Furthermore, PCA is commonly leveraged to extract dominant collective modes in simulations of proteins, referred to as ``Essential Dynamics'' in that field.~\cite{protein_pca_1,protein_pca_2} More recently, various spin models from statistical physics have been investigated via PCA and other machine learning methods.~\cite{spin_neural_classification,spin_confusion_ml,spin_ml_1, spin_ml_2,spin_ml_3,spin_ml_4,spin_ml_5,spin_ml_6} These studies have demonstrated the ability of machine learning tools to detect and quantify phase transitions by the autonomous construction of an order parameter (OP). 

The aforementioned work on phase transitions in spin models served as motivation for this two-part series of papers. In the first manuscript (henceforth referred to as Paper I), we developed guidelines for the utilization of PCA~\cite{br_and_ml_book,stat_learning_book,pca_guide} to detect phase transitions in off-lattice, particle-based systems. We also demonstrated that PCA can readily identify the freezing transitions in hard disks and hard spheres, as well as liquid-gas phase separation in a binary mixture of patchy particles with complementary attractions. In developing and evaluating this approach, we initially focused on phase transitions that were equilibrium in nature and could be identified on the basis of features reflecting the positional degrees of freedom of the particles.

Here, we seek to generalize the formalism developed in Paper I to assess its utility for detecting phase transitions in a broader class of systems. Examples include equilibrium systems with 1) anisotropic particles leading to orientational as well as positional ordering~\cite{liquid_crystals_1,liquid_crystals_2} and 2) compositional degrees of freedom that can induce demixing, even in the absence of appreciable density fluctuations.~\cite{general_phase_behavior} We also address active or driven matter, which exhibits phase transitions whose detection and characterization cannot generally be facilitated based on arguments from equilibrium statistical mechanics.~\cite{noneq_pts_review,noneq_pts_book_v1,noneq_pts_book_v2,active_matter_pts,oscill_shear_pts_1,oscill_shear_pts_2,active_particles_review,oscill_mag_assembly,driven_pts_ex,active_pts_ex} 

We propose several numerical encoding schemes (i.e., feature vector representations) for data describing particle configurations in these systems to detect their phase transitions with PCA. We find that prior knowledge of the phase transition is not required to construct a useful feature vector; consideration of the properties of the model system at hand is sufficient. However, we also show that by performing PCA on several choices for the feature vector, one can gain physical insights into the nature of the phase transition.

The balance of the manuscript is organized as follows. In Sect.~\ref{sec:methods}, considerations for constructing features for the detection of phase transitions in off-lattice systems using PCA are presented. The model systems analyzed in this work and the corresponding simulation details for each model are also provided. Sect.~\ref{sec:results} is divided into three subsections, each dedicated to a different model system. The first, Sect.~\ref{subsec:randorgR}, describes a study of the Random Organization Model, which exhibits a nonequilibrium phase transition between a quiescent state and a dynamically evolving steady state as a function of increasing density.~\cite{hyperuniformRO,OriginalRO,BerthierRO,SchmiedebergRO} Sect.~\ref{subsec:ellipsesR} addresses the fluid-nematic (orientationally driven) and the nematic-solid (positionally driven) phase transitions that occur upon densification of hard ellipses.~\cite{Baron_ellipses_k_6,ellipses_phase_diagram_Cuesta,ellipses_phase_diagram_Xu,ellipses_phase_diagram_Bautista} Finally, in Sect.~\ref{subsec:WRResult}, compositional demixing in the Widom-Rowlinson Model--a binary mixture where unlike particles interact via excluded volume effects but like particles are noninteracting\cite{WR,Ruelle,Chayes1995}--is explored. Concluding remarks are presented in Sect.~\ref{sec:conclusions}.     

\section{Methods} 
\label{sec:methods}
\subsection{Feature Construction}
\label{subsec:features}

Features ($f_i$) are scalar quantities that inform a machine learning algorithm about some aspect of the system being studied.~\cite{br_and_ml_book,stat_learning_book} Here, we denote a general vector of $m$ features as
\begin{equation} \label{eqn:general_features}
\begin{split}
&\boldsymbol{f} \equiv \begin{bmatrix}
    f_{1},      & f_{2}, & \dots,  & f_{m}     
\end{bmatrix}^{T},
\end{split}
\end{equation} 
where $T$ indicates a transpose. Feature vectors provide a numerical encoding for the separate realizations (or measurements) contained in the data set ($\mathcal{D}$). 

When possible, features should reflect any known constraints; for physics problems, these include invariance to translation and rotation.~\cite{md_1,md_2,schnet,multibody_expansion} Such constraints can be easily encoded via the use of internal coordinates (e.g., interparticle distances or relative angular orientations) as features. Here, we compute pairwise quantities $g_{\beta}^{(\alpha)}$ that are in reference to a probe particle ($\alpha$) and a corresponding particle in its environment ($\beta$). A feature vector built from information considering $n_{\text{P}}$ probe particles (each with $n_{\text{NN}}$ corresponding environmental particles) can be represented as
\begin{equation} \label{eqn:multi_probe_features}
\begin{split}
&\boldsymbol{f} = \begin{bmatrix}
    \boldsymbol{g}_{1}^{T},      & \boldsymbol{g}_{2}^{T}, & \dots,  & \boldsymbol{g}_{n_{\text{P}}}^{T}
\end{bmatrix}^{T}, \\
&\boldsymbol{g}_{\alpha}^{T} \equiv \begin{bmatrix}
g_{1}^{(\alpha)}, & g_{2}^{(\alpha)}, & \dots, & g_{n_{\text{NN}}}^{(\alpha)}
\end{bmatrix}
\end{split}
\end{equation} 
where the full vector of vectors $\boldsymbol{g}_{\alpha}^{T}$ corresponding to each probe particle $\alpha$ is implicitly flattened to form one contiguous feature vector (block matrix notation). 

Within the above mathematical framework, there is no unique choice for the selection of either the probe particles or the neighboring particles that define their environment. Once a collection of probe and corresponding environment particles are chosen, we also must specify how the resultant pairwise quantities (the $g_{\beta}^{(\alpha)}$) are assigned to the $\alpha$ and $\beta$ indices in Eq.~\ref{eqn:multi_probe_features}. So that we do not have to compute properties with respect to every particle in the simulation box, we select $n_{\text{P}}$ probe particles at random. For the corresponding environmental particles, we use physical intuition as a guide by assuming that the distance between the probe particle and a given environmental particle $r_{\beta}^{(\alpha)}$ will influence the manner in which the associated feature $g_{\beta}^{(\alpha)}$ reports on a given phase transition. As a result, we use a distance-based criterion to determine which particles comprise the environment for a given probe (e.g., the first twenty nearest neighbors or every tenth nearest neighbor), hence our use of $n_{\text{NN}}$ to denote the number of environmental particles. Similarly, we assign the index $\beta$ on the basis of interparticle distance so that   
\begin{equation} \label{eqn:nearest_neighbor_sorting}
r_{1}^{(\alpha)} \leq r_{2}^{(\alpha)} \leq \dots \leq r_{n_{\text{NN}}}^{(\alpha)}
\end{equation} 
 The assignment of a probe particle to a given $\alpha$ is less intuitive and could be model-dependent; however, random assignment is always a possibility, and, as we discuss below, the results obtained from that initial assignment can in some cases help to identify a superior assignment scheme for the probe particles. 

In principle, $n_{\text{P}}$ could be as large as the number of particles in the simulation, $N$, and $n_{\text{NN}}$ could have a maximum value of $N-1$. Since the total feature vector size (in relation to Eqn.~\ref{eqn:general_features}) is $m=n_{\text{P}}\times n_{\text{NN}}$, the preceding choices would yield a feature vector of length $N(N-1)$. For most systems of interest, PCA for feature vectors of this size would be computationally infeasible. Therefore, practical implementation of PCA using particle-based coordinate data requires sensible choices for $n_{\text{P}}$ and $n_{\text{NN}}$ that we describe in the following sub-sections.

Finally, we refer to features where the $g_{\beta}^{(\alpha)}$ are physically motivated quantities as ``intuited'' features ($\boldsymbol{f}_{\text{I}}$). In Paper I, we showed that $\boldsymbol{f}_{\text{I}}$ do not necessarily approximate white noise in the disordered reference state (here, the ideal gas) limit and therefore may possess correlations that could obscure the detection of a phase transition via PCA. Arriving at corrected features ($\boldsymbol{f}_{\text{C}}$) that are linearly decorrelated when applied to an ideal gas reference data set ($\mathcal{D}_{0}$) is accomplished by deriving a PCA whitening transformation~\cite{whitening} ($\boldsymbol{f}_{\text{I}}\rightarrow\boldsymbol{f}_{\text{C}}$) that satisfies $\langle \boldsymbol{f}_{\text{C}}\boldsymbol{f}_{\text{C}}^{T} \rangle_{\mathcal{D}_{0}} = \boldsymbol{I}$ where $\boldsymbol{I}$ is the unit matrix and $\langle \dots\rangle_{\mathcal{D}_{0}}$ is an average over the reference data. 

\subsection{Models}
\label{subsec:models}

We provide a brief description of each model examined in this work as well as the relevant phase transition(s) below. We then specify the form of the associated feature vectors within the framework provided by Eq.~\ref{eqn:multi_probe_features}. Finally, we describe the simulation protocols used to generate the configurations on which the PCA is performed. Throughout, $N$ denotes the number of particles in a two-dimensional (2D) periodically replicated simulation cell of area $A$, $\rho=N/A$ is the number density, and $\eta=\rho \pi \sigma^{2}/4$ is the packing fraction.

\subsubsection{Random Organization Model} 
\label{subsec:randorgM}

In one variant of the Random Organization (RandOrg) model, a circular particle of diameter $\sigma$ is defined as active if it overlaps with any other particle.~\cite{SchmiedebergRO,BerthierRO} For a given configuration, all active particles are simultaneously given random displacements; all other particle positions are unaltered. Particle positions are initialized at random, from which the above procedure is repeated until either 1) a so-called absorbing state is reached where no particle overlaps are present (lower densities) or 2) a steady-state is reached where the fraction of active particles fluctuates about some non-zero value (higher densities). 

Given that the RandOrg model comprises identical, radially symmetric particles, features that explicitly encode positional packing correlations around tagged particles are an obvious first choice to try. Specifically, we utilize mean subtracted interparticle distances as our features
\begin{equation} \label{eqn:distance_features}
g_{\beta}^{(\alpha)} = r_{\beta}^{(\alpha)} - \langle r_{\beta}^{(\alpha)} \rangle_{\mathcal{D}}
\end{equation}
Furthermore, while the model is technically single-component to the extent that there are no immutable labels associated with the particles, multiple particle types (active and inactive) are created on-the-fly due to the dynamics prescribed by the model. To capture emergent inhomogeneity with respect to particle environment, it is critical to utilize multiple probe particles as prescribed by Eq.~\ref{eqn:multi_probe_features}. In the present work, we use a fixed feature length of $m=n_{\text{P}} \times n_{\text{NN}}=400$, and examine the effects of co-varying $n_{\text{NN}}$ and $n_{\text{P}}$. 

Within the above approach, there is still the question of how to assign the probe particles to specific values of $\alpha$ in Eqn.~\ref{eqn:multi_probe_features}.
One valid, though perhaps not particularly informative, choice is to randomly order the probe particles, such that $\alpha$ assignment does not encode any information. In Sect.~\ref{subsec:randorgR}, we demonstrate how performing PCA with this choice produces results that suggest a more informative sorting scheme, where probe particles are assigned to the index $\alpha$ on the basis of their first NN distance, i.e. $r_{1}^{(1)} \leq r_{1}^{(2)} \leq \dots \leq r_{1}^{(n_{\text{P}})}$.

We note two additional technical points regarding PCA for the RandOrg model. First, as we increase $n_{\text{P}}$, the magnitude of the OP grows in a nonlinear fashion. We can collapse OPs onto the same scale by dividing by the square root of the explained variance of their dimension, a procedure equivalent to data ``whitening'' discussed in Paper I in the context of correcting the physics-motivated features. We also find that OPs obtained from both dimensional (preserving units of distance) and nondimensionalized features (dividing raw distances by $\rho^{-1/D}$, where $D$ is the dimension) accurately detect the phase transition of the RandOrg Model. However, as we demonstrate, the PCA-derived OP using the former convention shows behavior that is more strikingly reminiscent of the classical OP for this system.

To generate the configuration data required to construct the above features, we performed 2D simulations in a square box with $N=1000$ particles and employed a maximum displacement of 0.25$\sigma$ in both the $x$ and $y$ directions for the active particles. The length of the simulations varied with proximity to the critical point characteristic of the transition between an absorbing and a steady state. At densities below the critical point, an individual simulation ended when the absorbing state is reached; however, critical slowing down impacts the simulation length required to achieve that state.~\cite{BerthierRO,OriginalRO} We used a maximum of $10^5$ simulation steps for densities below the critical point. For the higher densities, the fraction of active particles decreased from the initial random state before fluctuating about a steady-state. The number of simulation steps was chosen to be at least twice as long as the initial relaxation time scale, ranging from $10^3$ steps (at the highest densities) to $10^5$ steps (just past the critical point). We performed $10^3$ separate simulations, using only the last frame from the simulation in the PCA. Values for $\rho$ ranging from $0.38$ to $0.63$ were simulated in increments of $0.005$. From the simulation data, we computed $25$ feature vectors from each simulation snapshot, where the probe particles were selected at random. Within a single feature vector, probe particles are selected without replacement; however, a particular probe particle can appear in multiple feature vectors. 

\subsubsection{Hard Ellipses}
\label{subsec:ellipsesM}

Densification of hard ellipses bears similarity to the freezing of hard disks studied in Paper I but with added complexity derived from particle-shape anisotropy. In addition to ordering on the center-of-mass positional level, quasi-long ranged orientational ordering is possible, yielding the so-called nematic phase.~\cite{liquid_crystals_1,liquid_crystals_2} Two obvious pairwise properties to compute from the configurational data of hard ellipses are center-of-mass distances and the relative orientations of the ellipses. With respect to the former, we use the positional features with a single probe particle as employed in Paper I for hard disks and spheres. This form is equivalent to the feature vector defined by
Eqs.~\ref{eqn:multi_probe_features}-\ref{eqn:distance_features} for the case where $n_{P}=1$ and $n_{\text{NN}}=N-1$, where $N$ is the number of particles. Subsequently, the size of the feature vector is reduced by only including every $10^{\text{th}}$ NN distance after the first feature in the final feature vector. The pairwise distances are normalized with respect to the mean interparticle spacing $l\equiv\rho^{-1/D}$, where $\rho$ is the number density and $D$ is the spatial dimensionality, to yield non-dimensionalized features. 

For the latter case of relative orientations, we still employ one probe and index its environmental particles on the basis of NN sorting (Eq.~\ref{eqn:multi_probe_features}-\ref{eqn:nearest_neighbor_sorting}); however, we use a measure of relative pair orientations in place of pair distances in the feature vector. Defining $\delta\theta_{\beta}^{(\alpha)}$ as the angular difference between the probe and environmental particles assigned to indices $\alpha$ and $\beta$ respectively, we employ features of the form
\begin{equation} \label{eqn:angular_features}
g_{\beta}^{(\alpha)}=\big|\cos{\big(\delta\theta_{\beta}^{(\alpha)}\big)}\big|-\big\langle\big|\cos{\big(\delta\theta_{\beta}^{(\alpha)}\big)}\big|\big\rangle_{\mathcal{D}}
\end{equation}
That is, $g_{1}^{(1)}$ quantifies the relative orientation of the single probe particle with its closest NN ellipse, etc. From the sorted list defined by the combination of Eq.~\ref{eqn:multi_probe_features}, Eq.~\ref{eqn:nearest_neighbor_sorting}, and Eq.~\ref{eqn:angular_features}, only every $10^{\text{th}}$ NN is included in the feature vector, as was done for the positional features above.

The feature vectors used as input to PCA were collected from Monte Carlo simulations of hard ellipses carried out at constant particle number and volume using the HOOMD-blue software package.~\cite{hoomd_1,hoomd_2,hoomd_3} The box shape was chosen to approximate a square by an appropriate distribution of triangular lattice cells with an aspect ratio of $\sqrt{3}\kappa$, where $\kappa=b/a$ is the ratio of the semi-major ($b$) and semi-minor ($a$) elliptical axes, respectively. (Here, we set the lengthscale as $2a=1$.) Specifically, given the number of cells in the $y$ direction, $n_y$, the number of cells in the $x$ direction is chosen as $n_x=\text{round}(\sqrt{3}\kappa)$. For ellipses with $\kappa=\{3, 4, 6, 9\}$, we chose $n_y$=\{17, 15, 12, 10\} which yielded total number of particles, $N$=\{2992, 3120, 3000, 3120\}, respectively. For each step, the move type (rotation or translation) was selected at random with equal probability. The maximum degree of translation and rotation per move were independently scaled to yield a $\sim25\%$ acceptance rate for efficient phase sampling. Density ranges were chosen to span the isotropic, nematic and solid phases. For ellipses with $\kappa=\{3, 4, 6, 9\}$, we chose $\eta=\{0.6-0.9, 0.55-0.9, 0.4-0.9, 0.3-0.9\}$, respectively. A typical run proceeded as follows. A system of $N$ hard ellipses was started from an ideal triangular lattice at maximum packing fraction and expanded to a target $\eta$ value. Next, the range of translational and rotation move sizes were optimized using 50 iterations of 100 steps to achieve the targeted acceptance ratio, where a step is equal to a HOOMD-blue ``timestep'', or approximately four sweeps over all particles. Then, the system was equilibrated for $6\times10^6$ steps, and data was collected every $6000$ steps from an additional $6\times10^6$ step production run. From each frame, 30 feature vectors were constructed, where the probe particles were selected without replacement within a given frame.

\subsubsection{Widom-Rowlinson Model} 
\label{subsec:WRModel}
The Widom-Rowlinson (WR) model~\cite{WR} is composed of a binary mixture of A and B particles where like pairs (A-A or B-B) are non-interacting and unlike pairs (A-B) interact isotropically via a hard-core repulsion with diameter~$\sigma$. Upon densification, the WR model compositionally demixes to form separate A- and B-rich phases. The resulting phase transition can straightforwardly be used to model compositional demixing; however, by integrating out the coordinates of one of the species, a model for liquid-gas coexistence can be obtained. In this work, we study the symmetric WR model where the number of A and B particles are equal.

Full specification of an individual particle in the WR model requires knowledge of both its type (A or B) and its position, yielding two obvious quantities to include in the feature construction. Instead of directly encoding the particle type as a categorical variable, we use particle type information to modify the assignments of the $\alpha$ and $\beta$ indices. We construct NN positional features as prescribed by (Eqs.~\ref{eqn:multi_probe_features}-\ref{eqn:distance_features}), but we only include distances between pairs of A particles in the feature vector. We use a single probe particle ($n_{\text{P}}=1$) with $n_{\text{NN}}=1200$ nearest neighbors for the environmental descriptors. By neglecting one of the WR components, we construct features that explicitly leverage both compositional and positional information. Finally, we non-dimensionalize the distances in the same fashion as the ellipse positional features described in Sect.~\ref{subsec:ellipsesM}.

For the production of the configuration data required to construct feature vectors for PCA, the HOOMD-Blue hard-particle Monte Carlo integrator~\cite{hoomd_1,hoomd_2,hoomd_3} was used to perform the simulations of the WR model in a square box for $N=4096$ particles in 2D. Equilibrium samples were generated at number densities spanning both the mixed and ordered phases as follows. After compressing the final configuration from simulation at the previous density, the system was equilibrated for $10^7$ steps, followed by a production run of $10^7$ steps, from which data was collected every $10^3$ steps, for a total of $10^4$ snapshots per density. A step is equivalent to a HOOMD-blue ``timestep'' as defined in Sect.~\ref{subsec:ellipsesM}. Simulations were run from $\rho = 0.064$ to $3.82$ in increments of $0.064$. From each frame, a single feature vector was constructed.

\section{Results and Discussion}
\label{sec:results}

Prior to examining the PCA results for the above models, we explain the general interpretation of the quantities that result from PCA below. For the features constructed according to the protocols described in Sect.~\ref{sec:methods}, PCA discovers a set of orthogonal axes--the principle components (PCs)--that are constructed in succession so as to maximize the data variance projected along each new axis. In this work, we monitor the relative explained variance of the PCs, denoted as $\lambda_{i}$ for the $i^{\text{th}}$ PC; by convention, the PCs are sorted so that $\lambda_{i}\ge \lambda_{i+1}$. A comparatively large value for $\lambda_1$ indicates that the information content of the features has been effectively concentrated into a single dimension: the first PC.

Of particular relevance to interpreting the PCA results is the projection of the feature vectors along the PCs: the PC score, denoted $p_{i}$ for the $i^{\text{th}}$ PC. Given that the first PC contains the largest explained variance, we evaluate the use of $p_{1}$ as an OP-like quantity to report on the phase transition of interest.~\footnote{We do not exclude the possibility that the other PC scores may be useful or that a better order parameter could involve multiple PC scores.~\cite{spin_ml_1,spin_ml_2,spin_ml_3,spin_ml_4,spin_ml_5,spin_ml_6} For simplicity we focus on $p_{1}$} This strategy amounts to coalescing as much ``information'' (i.e., variance) as possible from the high-dimensional feature vector $\boldsymbol{f}_{\text{C}}$ into the scalar $p_1$. Since each $p_1$ is associated with a single feature vector, we define two quantities that are averaged over a given state-point $\mathcal{S}$ (here, the state points are densities): $P_{1}\equiv \langle p_{1} \rangle_{\mathcal{S}}$ and the associated standard deviation, $\sigma_1\equiv\sqrt{\langle p_{1}^{2} \rangle_{\mathcal{S}}-\langle p_{1} \rangle_{\mathcal{S}}^{2}}$. When $P_1$ and $\sigma_1$ are plotted as a function of density, phase transitions will generally be indicated by a sigmoid in the former and a peak in latter metric.

The final relevant quantities from the PCA calculation are the PCs themselves--the weights that relate the features and the PC scores. As described in Sect.~\ref{subsec:features}, the intuited features ($\boldsymbol{f}_{\text{I}}$) are transformed in a corrected representation ($\boldsymbol{f}_{\text{C}}$), the latter of which are input into the PCA calculation. Since the values comprising $\boldsymbol{f}_{\text{I}}$ are straightforward to interpret physically, we explore the relationship between the PC scores and $\boldsymbol{f}_{\text{I}}$ (instead of $\boldsymbol{f}_{\text{C}}$). As described in Paper I, it is possible to write down a linear relationship between the scalar $p_{i}$ and the vector $\boldsymbol{f}_{\text{I}}$ via the following dot product
\begin{equation} \label{eqn:full_transformation}
{p}_{i} = \boldsymbol{q}_{i}^{T}\boldsymbol{f}_{\text{I}} \\
\end{equation}
where $\boldsymbol{q}_{i}$ is the desired vector of weights that map $\boldsymbol{f}_{\text{I}}$ to $p_i$. 
Examination of these weights reveals which physically meaningful quantities are particularly relevant to the phase transition.

In summary, we consider 1) the effectiveness of the dimensionality reduction via $\lambda_1$, 2) the low-dimensional (OP-like) representation of the data via quantities that depend on $p_1$ ($P_1$ and $\sigma_1$), and 3) the relative importance of the physical quantities that comprise $\boldsymbol{f}_{\text{I}}$ via the weights $\boldsymbol{q}_{i}$. For convenience, we summarize the above notation in the following table. 

\setlength{\arrayrulewidth}{1pt}
\setlength{\tabcolsep}{1pt}
\renewcommand{\arraystretch}{1.5}
\newcolumntype{C}{>{\centering\arraybackslash} m{0.8cm} }

\begin{table}
\caption{Common PCA variable definitions.}
\begin{center}
\label{tab:sometab}
\begin{tabular}{| C | m{7.2cm}|}
 \hline
 $\lambda_{i}$ & Relative (fractional) explained variance captured by the $i^{\text{th}}$ PC, ranging between 0 and 1, where larger values are indicative of greater importance.  \\
 \hline
 $p_{i}$ & The $i^{\text{th}}$ PC score. Mathematically, this is the projection of a feature vector along the $i^{\text{th}}$ PC. PCs offer a new coordinate system with information concentrated along the earlier (smaller index) PCs. \\
 \hline
 $P_{\text{i}}$ & Average of the $i^{\text{th}}$ PC score, $p_{i}$, over data from a state point ($\mathcal{S}$). $P_{1}$ serves as the OP-like quantity to report on phase transitions.\\
 \hline
 $\sigma_{i}$ & Standard deviation of the $i^{\text{th}}$ PC score, $p_{i}$, at a state point ($\mathcal{S}$). This is used as an effective ``susceptibility'' to locate the phase transition by identifying the maximum value. \\
 \hline
 $\boldsymbol{q}_{i}$ & Vector of weights that quantify the relevance of each feature to the $i^{\text{th}}$ PC.\\
 \hline
\end{tabular}
\end{center}
\end{table}

\subsection{Random Organization Model}
\label{subsec:randorgR}
The RandOrg model was developed to understand the transition from reversible to irreversible dynamics that occurs upon increasing either the applied periodic shear or the density of a material.~\cite{OriginalRO} In the first incarnation of the RandOrg model, an initial configuration was sheared and any particles overlapping with others as a result of deforming the simulation box were defined as active particles. Only the active particles were given a random displacement after which the simulation box was restored to its original geometry. At sufficiently low combinations of density and applied shear, a quiescent ``absorbing'' state eventually results, where shear does not generate further particle overlaps and there are no longer active particles. However, at greater densities and/or shear rates, shearing the system will always generate some overlaps. The reversible-to-irreversible transition reflects a state where the onset of particle collisions upon shearing prevents the system from returning to it original state when the shear is reversed.~\cite{OriginalRO,hyperuniformRO}

\begin{figure}
  \includegraphics{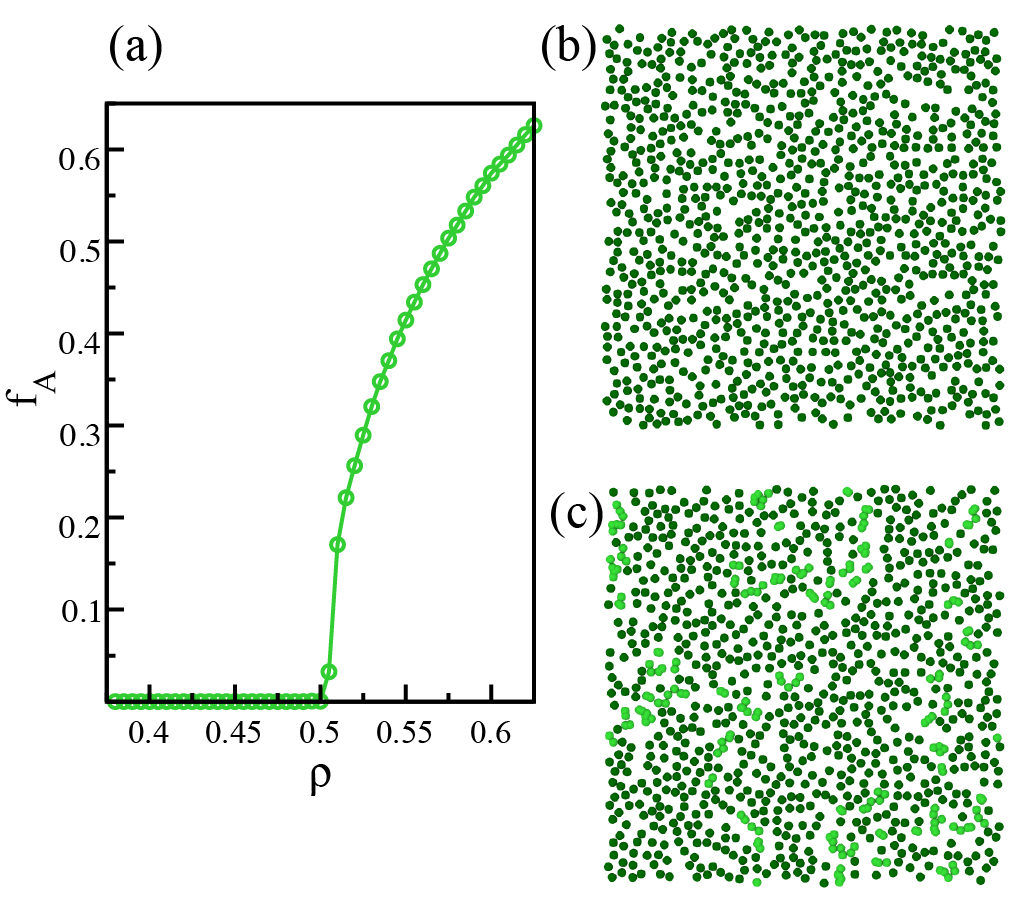}
  \caption{(a) Fraction of active particles, $f_{\text{A}}$, in the RandOrg model as a function of number density, $\rho$. (b,c) Simulation snapshots (b) below ($\rho=0.5$) and (c) above ($\rho=0.51$) the critical density. Active particles are shown in lighter green in panel c.}
  \label{fgr:RO_transition}
\end{figure}

A modified version of the RandOrg model, where shear is not included, has also been studied.~\cite{BerthierRO,SchmiedebergRO} Instead, as described in Sect.~\ref{subsec:randorgM}, initial particles are placed at random; active particles correspond to overlapping particles. This model possesses the same type of transition from an absorbing state at sufficiently low densities to an evolving steady-state containing a non-zero number of active particles at higher densities, while being technically simpler to implement. Fig.~\ref{fgr:RO_transition}a shows the fraction of active particles $f_{\text{A}}$ in this version of the RandOrg model as a function of number density $\rho$. Two simulation configurations, below ($\rho=0.5$) and above ($\rho=0.51$), the critical point are shown in Fig.~\ref{fgr:RO_transition}b and c, respectively.

Like hard disks, the RandOrg model comprises identical, radially symmetric particles, for which distance-based features are a sensible choice. Therefore, we first employed the feature vector developed in Paper I for hard disks--sorted nearest-neighbor (NN) distances associated with a single probe particle (Eqns.~\ref{eqn:multi_probe_features}-\ref{eqn:distance_features} where $n_{\text{P}}=1$). However, this construction of the feature vector did not produce a satisfactory OP. Use of a feature vector for which $n_{\text{P}}=1$ likely fails because, above the critical point, the system always has two effective particle types--active and inactive (see Fig.~\ref{fgr:RO_transition}c). Therefore, the environment of a single particle is not an accurate representation of the simulation box as a whole at higher densities.

\begin{figure}
  \includegraphics{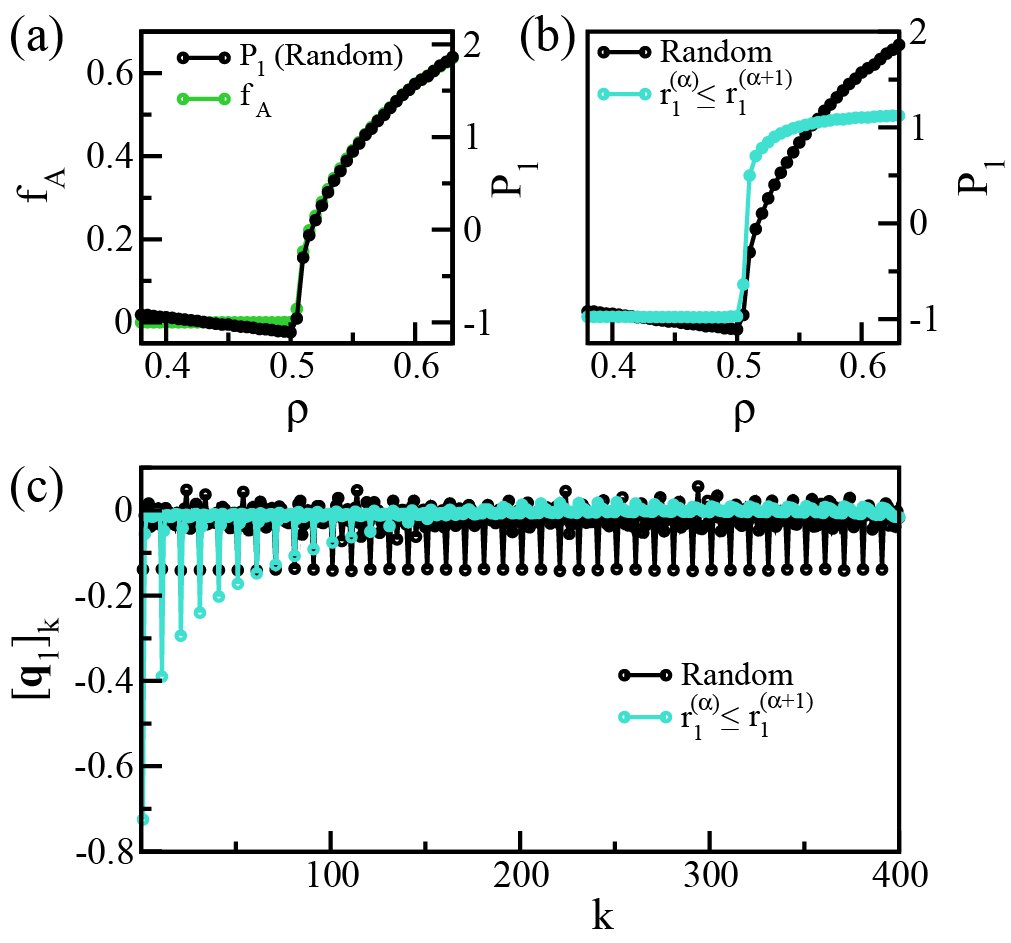} 
  \caption{(a) The PCA-deduced OP $P_{1}$ (with probe particles sorted randomly in the feature vector) as a function of number density $\rho$ compared to the conventional OP (the fraction of active particles, $f_{A}$) for the RandOrg model. (b) Comparison of $P_{1}$ with probe particles sorted randomly (black) versus according to their first NN distance so that $r_{1}^{(\alpha)} \leq r_{1}^{(\alpha+1)}$ (blue). (c) Component weights, $[\boldsymbol{q}_{1}]_{k}$, as a function of feature dimension $k$ for the two PC scores shown in panel (b).}
  \label{fgr:RO_random}
\end{figure}

As described in Sect.~\ref{subsec:features}, distance-based feature vectors can naturally incorporate multiple probe particles (and their corresponding neighbors). As with the NN distances for a given probe particle, one must decide how to order the probe particles inside the feature vector. In the absence of any information about the nature of a given phase transition, a first choice might be to randomly order the probe particles. In Fig.~\ref{fgr:RO_random}, we show results for the first PC using a feature vector constructed from 40 probe particles ($n_{\text{P}}=40$), each of which is encoded via its first 10 NN distances ($n_{\text{NN}}=10$). In principle, the weights associated with each probe particle should be identical since there is no physics-based interpretation for their ordering in the feature vector. Indeed in Fig.~\ref{fgr:RO_random}c we find a repeating pattern for every 10 weights: the first NN distance ($r_{1}^{(\alpha)}$) component weight for each probe is large in magnitude. The relative uniformity of the first NN distance weights, compared to the noisiness in the larger NN distances, indicates that the $r_{1}^{(\alpha)}$ values are informative to the PCA.

The corresponding OP is shown in Fig.~\ref{fgr:RO_random}a and bears striking resemblance to the standard OP shown in Fig.~\ref{fgr:RO_transition}a. Indeed, by arbitrarily shifting and scaling the PC score, we find that $f_{\text{A}}$ essentially overlaps with the PCA-deduced OP. It seems that the repeating unit in the component weights is able to distinguish between overlapping and non-overlapping particles and therefore can report on the relative amounts of active and inactive particles at a given value of $\rho$.

We can use the above component weights to intelligently devise a better sorting scheme for the probe particles. From Fig.~\ref{fgr:RO_random}c, it is clear that $r_{1}^{(\alpha)}$ is a highly weighted contribution to the feature vector; therefore, we performed a separate PCA calculation with the same values for $n_{\text{P}}$ and $n_{\text{NN}}$ while sorting the probe particles so that $r_{1}^{(\alpha)} \leq r_{1}^{(\alpha+1)}$. Because we have sorted the probe particles on the basis of a physically meaningful descriptor, the symmetry among probe particles is broken and the probe particles with the closest NNs (i.e., those assigned to lower $\alpha$) are weighted more heavily than other probe particles (Fig.~\ref{fgr:RO_random}c). Moreover, the associated OP is significantly sharper at the phase transition, essentially giving a binary classification into absorbing states and dynamic steady states on the basis of the OP (Fig.~\ref{fgr:RO_random}b).

\begin{figure}
  \includegraphics{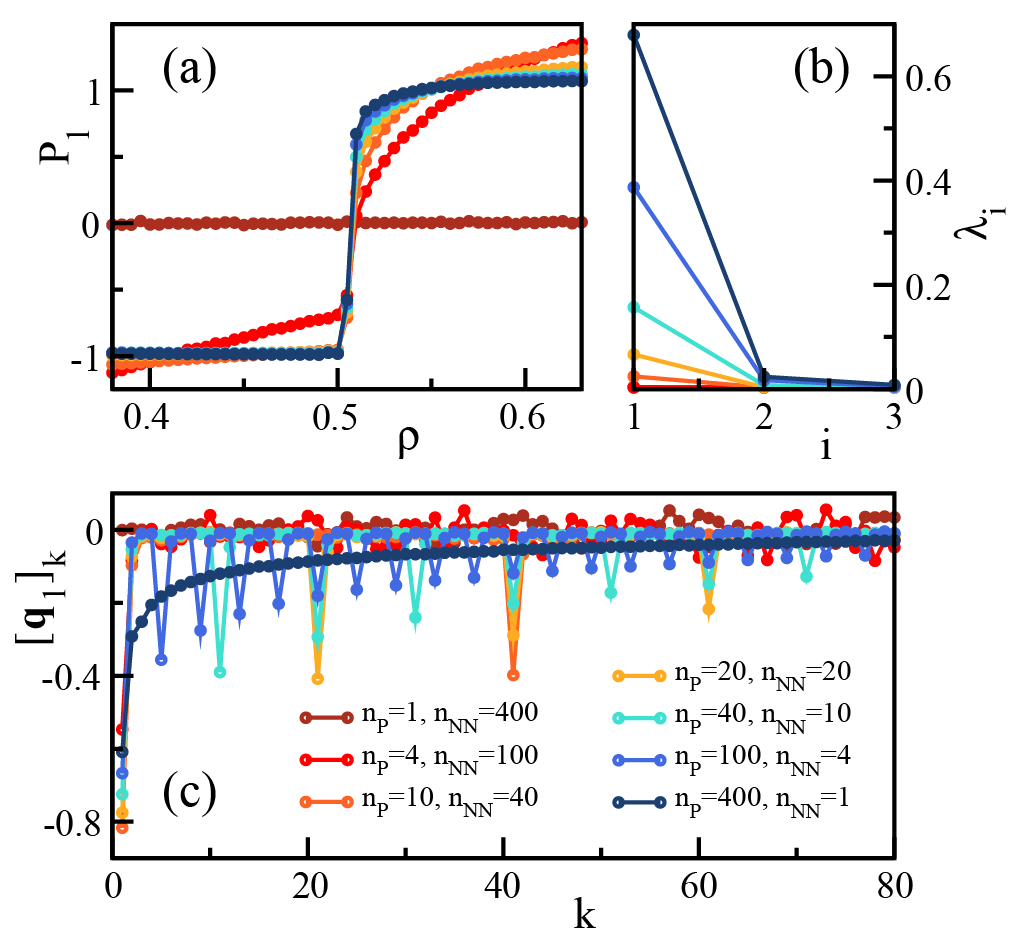}
  \caption{(a) PCA-deduced OP $P_{1}$ of the RandOrg model as a function of number density $\rho$ for different numbers of probe particles $n_{\text{P}}$ and corresponding nearest neighbors $n_{\text{NN}}$, respectively. (b) Corresponding explained variance for the first three PCs and (c) the first 80 PC weights $[\boldsymbol{q}_{1}]_{k}$.}
  \label{fgr:RO_PCA}
\end{figure}

With the above sorting scheme in hand, we vary both $n_{\text{P}}$ and $n_{\text{NN}}$ while keeping the length of the feature vector fixed at $m=n_{\text{P}} \times n_{\text{NN}}=400$. As $n_{\text{P}}$ increases and therefore $n_{\text{NN}}$ decreases, the quality of the first PC score as an OP improves significantly, with the metric sharpening into a sigmoidal curve that separates quiescent absorbing states from diffusive steady-states; see Fig.~\ref{fgr:RO_PCA}a. Conversely when $n_{\text{P}}=1$ (as was the case for Paper I), $P_{1}$ cannot detect the transition. Correspondingly, when features constructed with more probe particles are used, the explained variance associated with the first PC increases dramatically (Fig.~\ref{fgr:RO_PCA}b). The preceding trend is monotonic--there is no value to including more than the nearest interparticle distance per probe particle at constant $m$, an indication of the local character of the phase transition in the RandOrg model. 

For the above series of PCA calculations, the first 80 component weights are plotted in Fig.~\ref{fgr:RO_PCA}c. When $n_{\text{P}}$ is small, the components appear to be largely random, but as $n_{\text{P}}$ is increased, the components develop more structuring. For each probe particle, the component weights associated with $r_{1}^{(\alpha)}$ have a much larger weight than that of the rest of the features, reinforcing the importance of the first NN distance in the dimensionality reduction. Moreover, the probe particles that have closer first NNs have greater weights; we can interpret the role of using multiple probe particles as capturing an accurate measure of $r_{1}^{(1)}$ in a statistical sense, i.e., not all probe particles are required, but sampling is needed to make sure that sufficiently representative interparticle separations are included in each feature vector. 

While the importance of the first NN distance is intuitive given that the RandOrg phase transition is defined by the presence or absence of particle overlaps, we did not incorporate knowledge of the transition in constructing the features. In other words, our results suggest that modifying the feature vector can be used to infer characteristics of a transition, even if its nature is unknown at the outset. Specifically, for the RandOrg model, the importance of the first NN distance revealed by the PCA implies a transition that is local in character in real-space, and the necessity of multiple probe particles indicates that multiple distinct particle types or environments are an important characteristic.

\subsection{Hard Ellipses} 
\label{subsec:ellipsesR}

\begin{figure}[ht]
\includegraphics{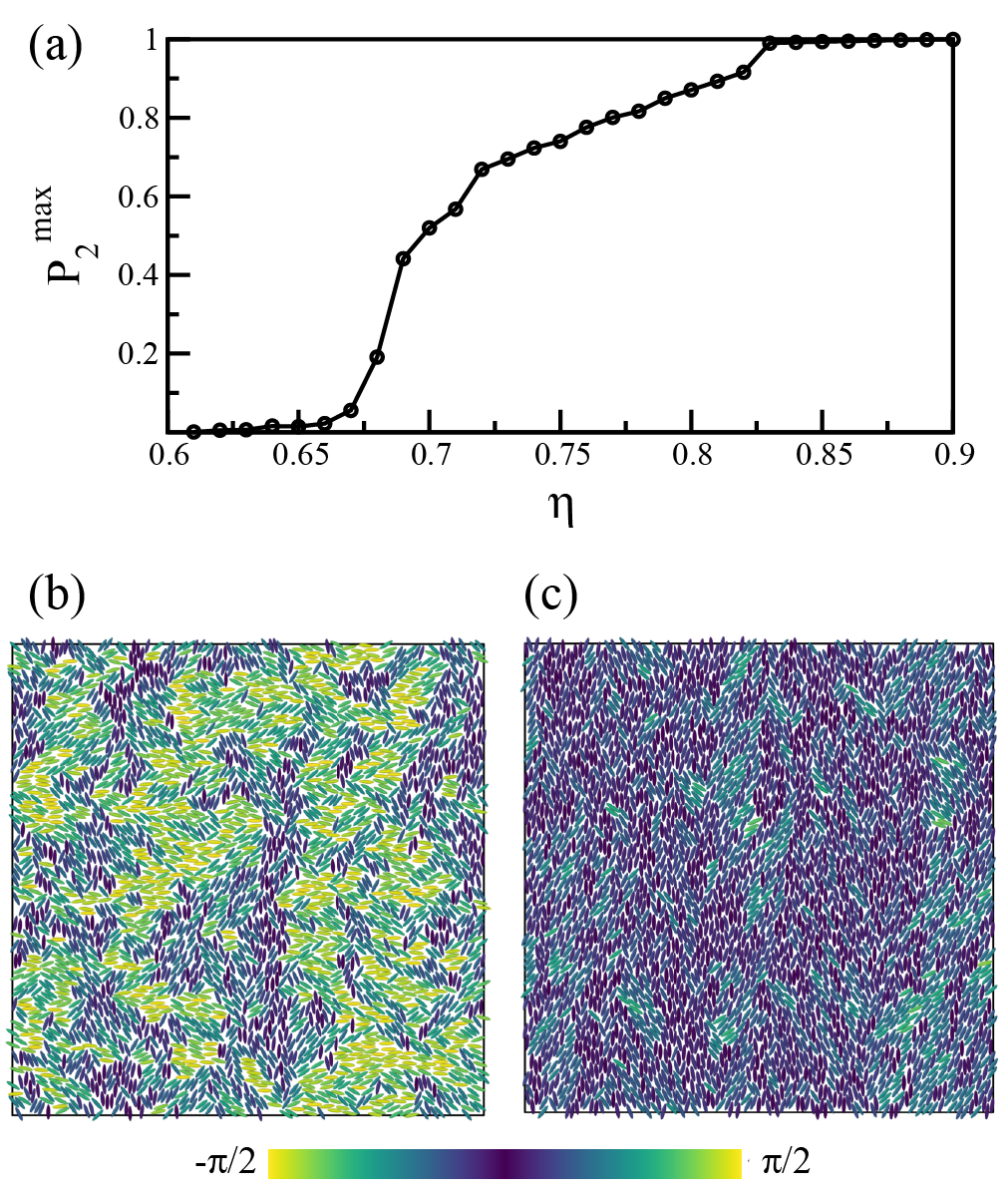} 
\caption{Density-driven isotropic fluid to nematic phase transition in a system of hard ellipses with aspect ratio $\kappa=4$. (a) The packing fraction $\eta$ dependence of the conventional order parameter for this transition, $P^{\text{max}}_{2}=[\langle 1/N \Sigma_{i}^{N} \text{cos}(2\theta_{i})\rangle ^{2}+\langle 1/N \Sigma_{i}^{N} \text{sin}(2\theta_{i})\rangle ^{2}]^{1/2}$, where $\theta_{i}$ is the angle between the semi-minor axis of the $i^{\text{th}}$ ellipse and the x-axis and $N$ is the number of ellipses, as per Ref.~\citenum{ellipses_phase_diagram_Bautista}. Simulation configurations of (b) the isotropic fluid at $\eta=0.65$ and (c) the nematic phase at $\eta=0.75$. Ellipses are color coded according to $\theta_{i}$ as defined above with the angular range limited to $[-\pi/2,\pi/2]$ due to orientation symmetry of the ellipse.} 	 
\label{fig:system_illustration}
\end{figure} 

The freezing transition for hard ellipses differs from that of hard disks because the former features an intervening nematic phase between the disordered fluid and the positionally ordered solid. The nematic phase manifests when the ellipses display disordered center-of-mass positions but quasi-long range orientational order.~\cite{Baron_ellipses_k_6,ellipses_phase_diagram_Cuesta,ellipses_phase_diagram_Bautista,ellipses_phase_diagram_Xu}. A conventional OP that reports on the the fluid-nematic transition, $P_2^{\text{max}}$, as well as simulation configurations at densities below and above the phase transition, are shown in Fig.~\ref{fig:system_illustration}. The continuous, second-order nature of the fluid-nematic transition is apparent from the behavior of $P_2^{\text{max}}$, from which the precise density for the underlying phase transition is not readily apparent. Therefore, one typically monitors the long-range power-law decay of a pairwise angular correlation function versus interparticle separation to identify the transition. The nematic phase transition point is identified when the power law decay exponent falls below an approximate value of $\frac{1}{4}$.~\cite{ellipses_phase_diagram_Cuesta,ellipses_phase_diagram_Xu}

\begin{figure}[ht]
\includegraphics{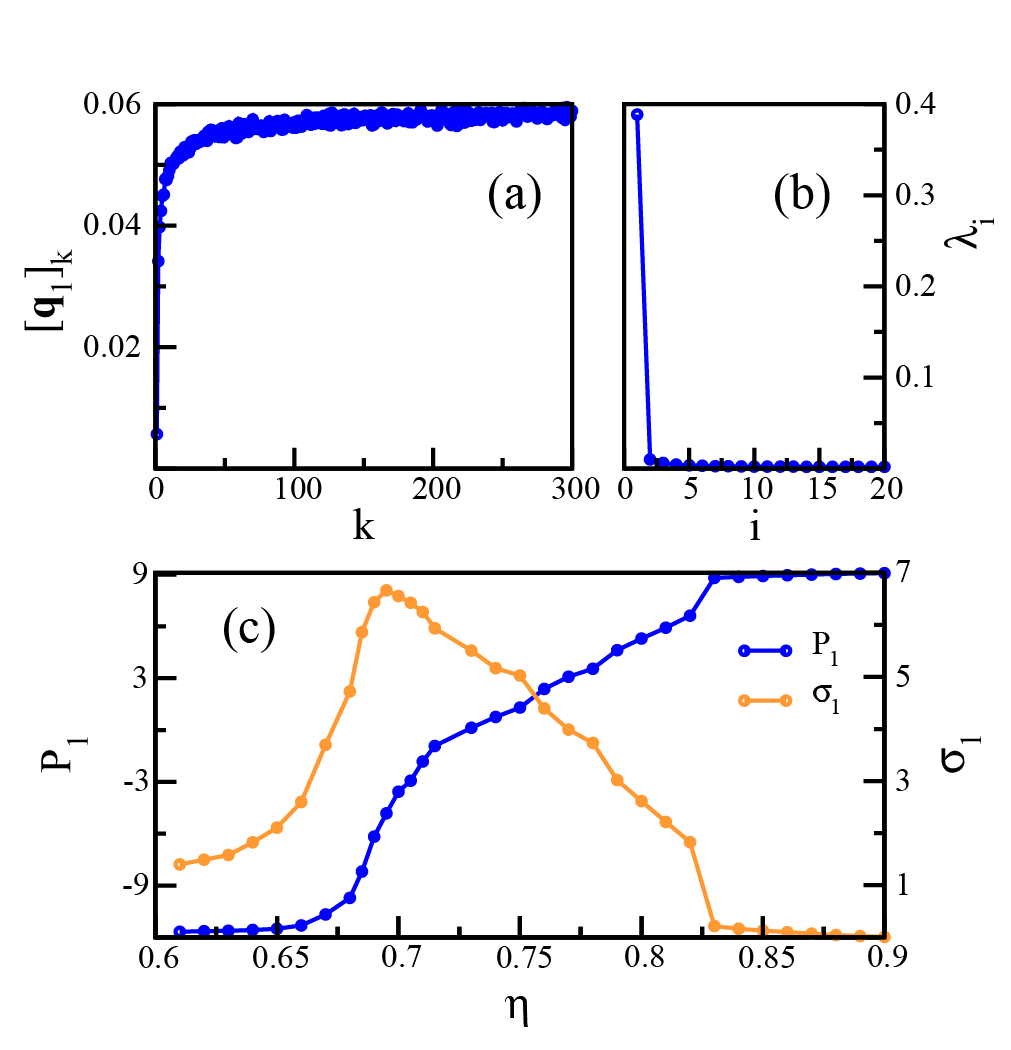} 
\caption{Based on PCA of the 2D system of hard ellipses, we show (a) component weights $[\mathbf{q}_{1}]_{k}$, (b) the explained variance $\lambda_{i}$, and (c) the OP ($P_1$) and standard deviation, $\sigma_1=\sqrt{\left \langle p_1^2 \right \rangle_{\mathcal{S}} -\left \langle p_1 \right \rangle_{\mathcal{S}}^2}$, where $p_1$ is the first PC score associated with an individual feature vector. Averages are taken over all feature vectors at a given state point ($\mathcal{S}$)--here, a single density.}
\label{fig:OPprops}
\end{figure} 

To detect the fluid-nematic phase transition via PCA, we use a feature vector constructed from the relative orientation of pairs of ellipses that are sorted in ascending order by the distance between the probe particle and its neighbor as described in Sect.~\ref{subsec:ellipsesM}. In Fig.~\ref{fig:OPprops}, we present the results of PCA using this orientational feature vector for ellipses with an aspect ratio $\kappa=4$. As seen by the explained variance $\lambda_i$ in Fig.~\ref{fig:OPprops}b, the first PC captures approximately $40\%$ of the data variance, indicating effective dimensionality reduction. From the component weights $[\mathbf{q}_{1}]_{k}$ in Fig.~\ref{fig:OPprops}a, it is clear that long-range orientations (larger values of $k$) are much more important than the closer neighbor orientations which tend towards zero. 
	
The above weights reflect the underlying structural motifs present in hard ellipses at various values of $\eta$. Orientationally aligned clusters of ellipses are present in both fluid and nematic phases (compare, for instance, the snapshots in Fig.~\ref{fig:system_illustration}b,c). Therefore, orientations between nearby ellipses (smaller values of $k$ in Fig.~\ref{fig:OPprops}a) are not useful indicators of nascent orientational long-range order, and their contributions to the OP are suppressed by the PCA. On the other hand, long distance components are approximately equal-weighed as they correlate proportionally to the presence of an emerging nematic director but average out for random configurations in the fluid state.
	
\begin{figure}[ht]
\includegraphics{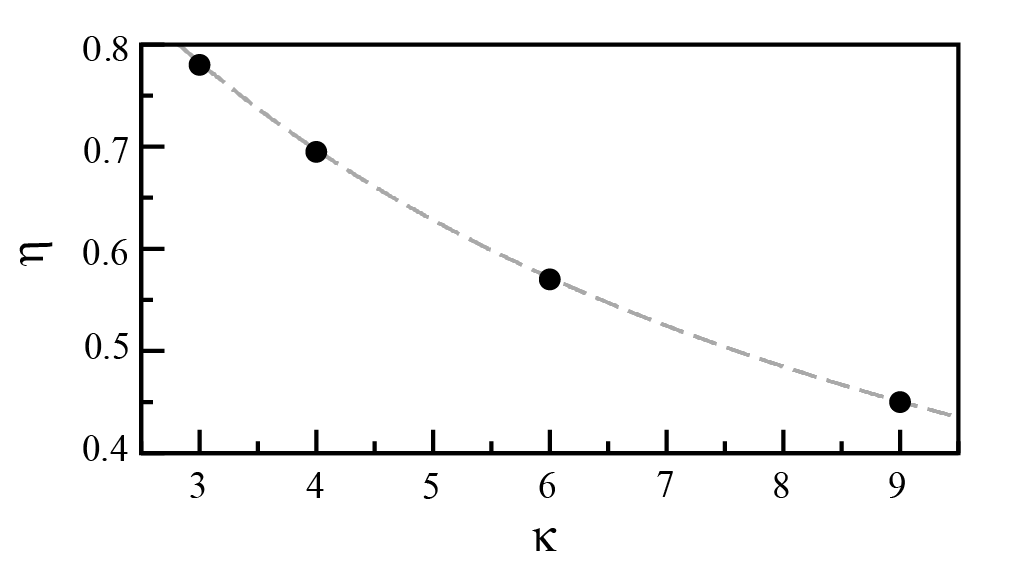} 
\caption{Phase boundary of the isotropic fluid to nematic transition for a system of hard ellipses as a function of packing fraction $\eta$ and aspect ratio $\kappa$. Solid black dots indicate the phase transition point identified from the position of maximum susceptibility, $\text{max}(\sigma_1)$. The dashed gray line indicates the phase boundary fit reported by Xu et al.\cite{ellipses_phase_diagram_Xu}}
\label{fig:nematic_boundary}
\end{figure} 
	
Regarding the OP itself, we show $P_1$ and its standard deviation $\sigma_1$ in Fig.~\ref{fig:OPprops}c; the latter quantity can be interpreted as a type of susceptibility of the OP. Note that $P_1$ resembles the traditional OP $P_2^{\text{max}}$ of Fig.~\ref{fig:system_illustration} and, while responsive to the nematic phase change, does not provide a unique transition point. As such, we use $\sigma_1$ to correlate the PCA results to the fluid-nematic phase transition: the maximum of $\sigma_1$ indicates the region most consistent with large-scale configuration fluctuations near the critical point of a continuous phase transition. Indeed, for all values of $\kappa$ investigated here, we find that the density $\eta$ associated with the maximum value of $\sigma_{1}$ is in excellent agreement with fitted fluid-nematic boundary reported by Xu et al\cite{ellipses_phase_diagram_Xu} (see Fig.~\ref{fig:nematic_boundary}), without requiring a tedious analysis of the long range scaling behavior in the angular correlations employed by the latter study.
 
\begin{figure}[ht]
\includegraphics{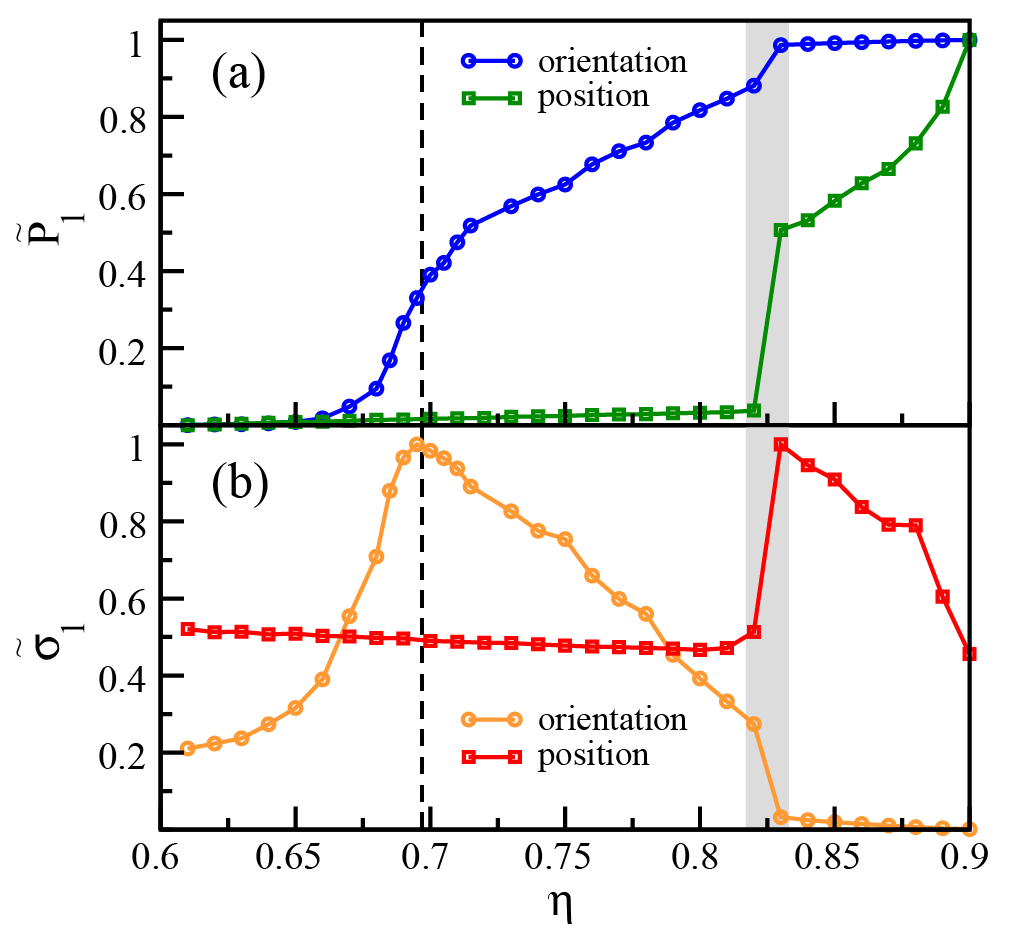} 
\caption{For the first PC of hard ellipses, comparison of the (a) shifted and normalized PC scores $\tilde{P}_1 \equiv \frac{P_1-\text{min}(P_1)}{\text{max}(P_1-\text{min}(P_1))}$ and (b) normalized standard deviations $\tilde{\sigma}_1 \equiv \frac{\sigma_{1}}{\text{max}(\sigma_{1})}$, as derived from either orientational or positional feature vectors. The dashed black vertical line indicates fluid-nematic boundary, and the shaded gray region indicates the nematic-solid phase coexistence region reported in Ref.~\citenum{ellipses_phase_diagram_Bautista}.}
\label{fig:pos_vs_orientation}
\end{figure} 
	
While use of orientational features as input to PCA provides a means to detect the fluid-nematic phase transition, the relationship between the above PCA results and the nematic-solid transition is less obvious. In Fig.~\ref{fig:pos_vs_orientation}a,b, we plot a normalized version of $P_1$ and its standard deviation $\sigma_{1}$ derived from the orientational features against the known nematic-solid coexistence region (the gray shaded area). There is a weak response to the nematic-solid region as $\tilde{\sigma}_{1}$ drops abruptly--perhaps an indicator of reduced orientational freedom of the ellipses upon solidification. However, the relationship between the angular degrees of freedom and the nematic-solid transition is relatively indirect and therefore performing PCA on the orientational features yields comparatively poor OP-like quantities for this phase change.
	
In order to detect the center-of-mass level ordering that occurs at the nematic-solid transition, we employ the positional NN features used for hard disks in Paper I. That is, instead of including the relative angle between two ellipses in the feature vector, we employ the interparticle distances. In Fig.~\ref{fig:pos_vs_orientation}a,b, we compare the normalized PC scores $\tilde{P}_1$ and the associated $\tilde{\sigma}_{1}$, respectively, when these positional features are used as input to the PCA. The resulting OP is insensitive to the fluid-nematic boundary but grows sharply across the known nematic-solid phase-coexistence region. Similar to the orientational features, the maximum in the position-based susceptibility $\sigma_1$ in Fig.~\ref{fig:pos_vs_orientation}b is an appropriate identifier for the underlying phase transition. The form of the OP as a function of $\eta$--flat over the fluid phase, rapid growth upon solidification, and more muted growth in the solid phase--is qualitatively similar to the OP reported in Paper I for the densification of hard disks. Together, the above results attest to the ability of PCA to provide insights into the character of a given phase transition by varying the form of the feature vector.  

\subsection{Widom-Rowlinson Model}
 
\label{subsec:WRResult}

\begin{figure}
  \includegraphics[width=3.37in,keepaspectratio]{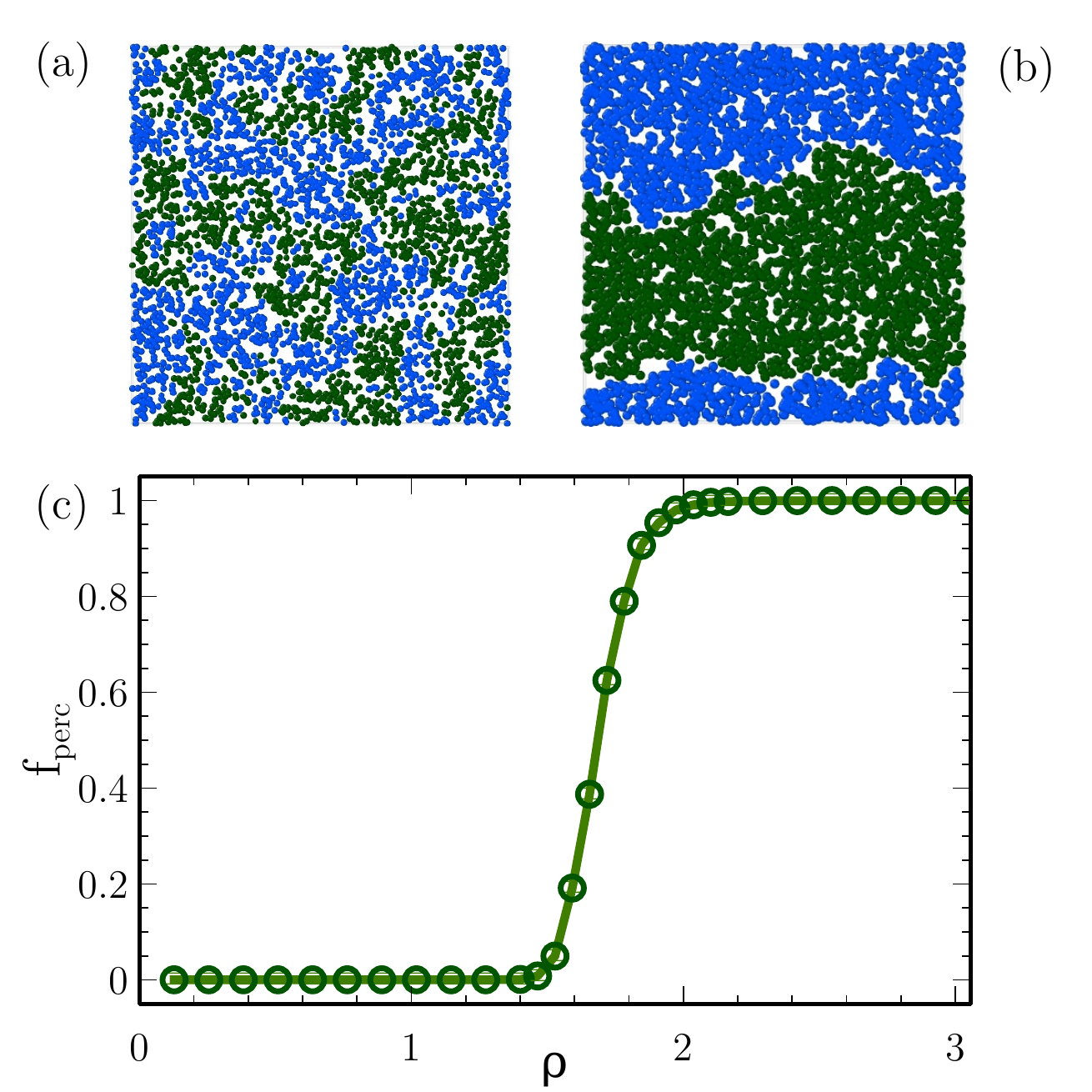} 
  \caption{For the WR model at $x=0.5$, simulated configuration snapshots of (a) mixed WR particles at $\rho = 1.25$ (below the phase transition) and (b) demixed WR particles at $\rho = 2.5$ (above the phase transition).(c) Fraction of percolated configurations ($f_{\text{perc}}$) as a function of number density.}
  \label{fig:TraditionalOP}
\end{figure}

As mentioned in Sect.~\ref{subsec:WRModel}, the WR Model contains two particles types--A and B--where like particles are non-interacting and unlike particles interact via a hard-core repulsion of diameter $\sigma$. At low densities, the two species are mixed. However, upon densification, a phase transition occurs~\cite{Ruelle,Chayes1995,Georgii} where the WR mixture phase separates into A-rich and B-rich regions (Fig.~\ref{fig:TraditionalOP}a,b) as the excluded volume effects experienced by the unlike particles overcomes the mixing entropy. The density at which the demixing transition occurs varies with composition; we denote $x$ as the fraction of A particles. In the present work, we study the mixture for which $x=0.5$.

When $x=0.5$, the density at which clusters of like particles become percolated can be used to determine the demixing transition.~\cite{Chayes1995,Klein,ChayesPRE} For the WR model, a cluster is defined as a group of particles that are all either directly overlapping or connected via a contiguous pathway of overlapping particles when periodic boundary conditions are properly taken into account. For a finite-sized, periodically replicated simulation box, a percolated cluster is one that grows in size upon replication of the simulation cell. Therefore, for each species at $x=0.5$, we computed the fraction of configurations possessing at least one percolated cluster of that particle type and averaged the results for the A and B particles to yield $f_{\text{perc}}$. Fig.~\ref{fig:TraditionalOP}c shows $f_{\text{perc}}$ as a function of density; percolated clusters were identified as described in Ref.~\citenum{linkergel}. One choice for the percolation threshold--the point when at least 50\% of the configurations are percolated--yields a de-mixing transition density of $\rho_{t}$ = 1.68. 

\begin{figure}
  \includegraphics[width=3.37in,keepaspectratio]{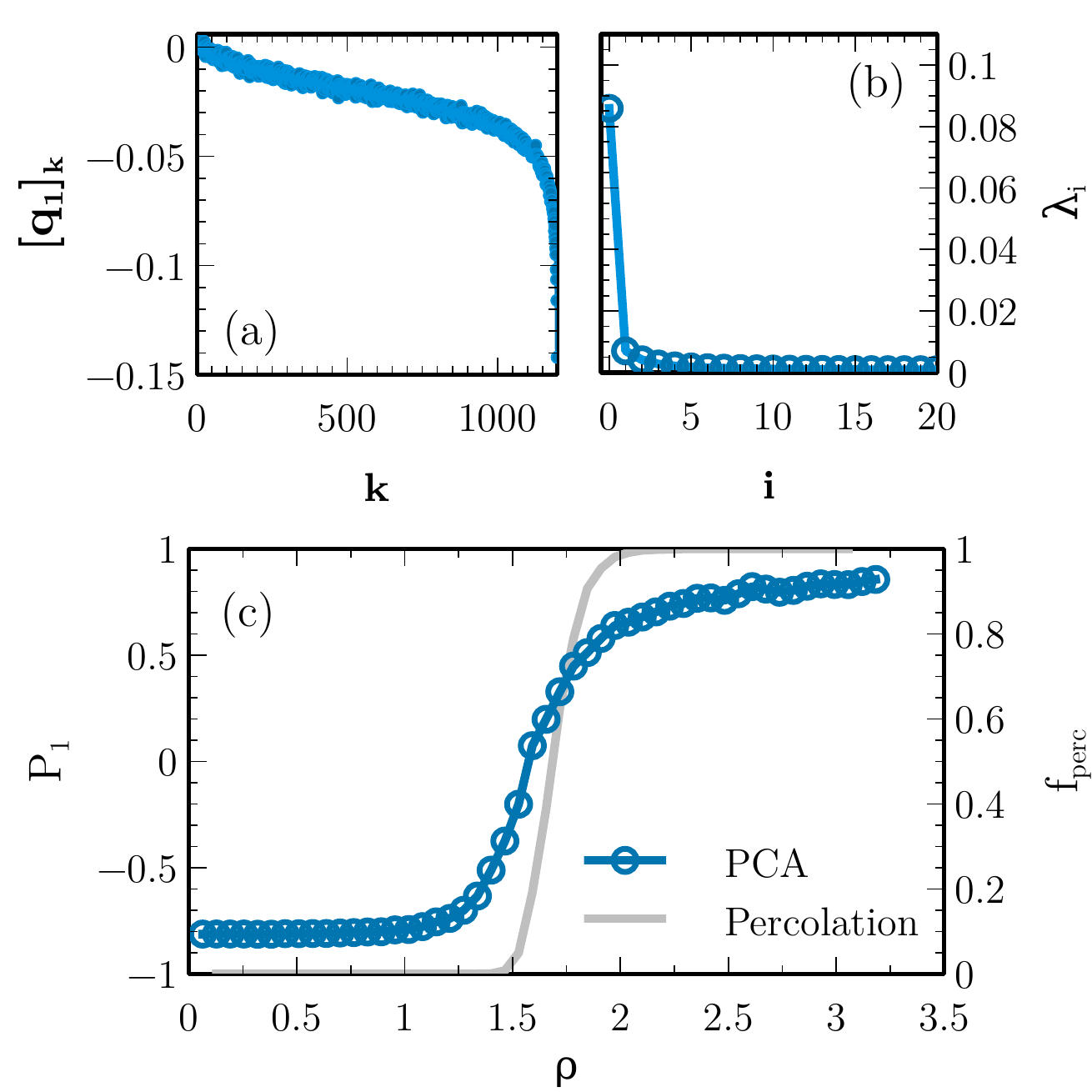}
  \caption{For the first PC upon application of PCA to the WR model, (a) its component weights $[\boldsymbol{q}_{1}]_{k}$, (b) the explained variance $\lambda_{i}$, and (c) the PCA-deduced OP ($P_{1}$) and percolation-based OP ($f_{\text{perc}}$).}
  \label{fig:LGFig}
\end{figure}

Positional features of the form defined by Eq.~\ref{eqn:multi_probe_features}-~\ref{eqn:distance_features} are unable to detect the above de-mixing transition if compositional degrees of freedom are not taken into account--a consequence of the absence of large scale fluctuations in the packings of the particles (agnostic to particle type) as the phase transition occurs. However, for any multicomponent mixture, features can be constructed using particle type data as well as spatial information. For the WR model, one such strategy is to design a feature vector that only includes interparticle distances if the corresponding pair of particles meets some criterion based on particle type. One such choice (though others are possible) is to only include distances between two A particles in the feature vector defined by Eq.~\ref{eqn:multi_probe_features}-~\ref{eqn:distance_features}--akin to the liquid-gas formulation of the WR model. The outcome of PCA with the above feature vector is shown in Fig.~\ref{fig:LGFig}. The component weights $[\boldsymbol{q}_{1}]_{k}$ in Fig.~\ref{fig:LGFig}a indicate that the long-ranged positional correlations dominate, whereas the smallest interparticle separations with respect to a given probe are essentially meaningless to the PCA. Reminiscent of the fluid-nematic transition seen in ellipses and described in Sect.~\ref{subsec:ellipsesR}, some local clustering on the basis of particle type occurs at lower densities than phase separation does; see Fig.~\ref{fig:TraditionalOP}a for example. Therefore, it is the long-range correlations that change sharply as the phase transition occurs. Fig.~\ref{fig:LGFig}b depicts the explained variance $\lambda_{i}$ for the first 20 PCs, where the first PC accounts for $\sim10\%$ of the data variance--an order of magnitude more than the succeeding components.

The PCA-deduced OP is shown in Fig.~\ref{fig:LGFig}c. Relative to $f_{\text{perc}}$, the PCA-based OP varies more slowly and has a wider transition window. To identify the unique transition point, the standard deviation $\sigma_{1}$, as outlined in Sect.~\ref{subsec:ellipsesR}, is computed for every density. The maximum value of $\sigma_{1}$, denoting the region with most variance in the feature vectors, is accepted as being associated with the transition density $\rho_{t}$. For the above one-component mixture, a sharp peak (not shown) in standard deviation is observed at $\rho_t$ = 1.66 which is in excellent agreement with the value obtained through percolation arguments ($\rho_{t}$ = 1.68), indicating successful identification of the de-mixing transition.

\section{Conclusions}


In this article, we extended the PCA framework introduced in Paper I to detect phase transitions in three new model systems, each characterized by very different physics. The success of the method in these cases highlights the importance of exploring various feature representations when seeking to detect such phase transitions and understand their underlying physics. 

Moving forward, two avenues seem fruitful for developing a routine analysis toolkit: (1) curate a sufficiently diverse library of features, each focused on different physical aspects that may be relevant to various phase transitions of interest and (2) explore the ability of more sophisticated, nonlinear learners to autonomously extract the physical intuition underlying such transitions on-the-fly. 

Finally, we comment on general trends that we observed to indicate that the PCA calculation was usefully reporting on the phase transition of interest. 
First we note that meaningful dimensionality reduction into the first PC is generally indicated by a large relative explained variance in comparison to the higher order PCs. Furthermore, we found that appropriate choices for the features resulted in an OP with strong convergence properties that required relatively small amounts of data to overcome sampling noise. We expect that these trends are relevant to other machine-learning approaches for the detection of phase transitions as well. 

\label{sec:conclusions}

\section*{Acknowledgments}
The authors thank Michael P. Howard for valuable discussions and feedback. This research was primarily supported by the National Science Foundation through the Center for Dynamics and Control of Materials: an NSF MRSEC under Cooperative Agreement No. DMR-1720595 as well as the Welch Foundation (F-1696). We acknowledge the Texas Advanced Computing Center (TACC) at The University of Texas at Austin for providing HPC resources.

\setcounter{figure}{0}
\setcounter{equation}{0}
\renewcommand\thefigure{A\arabic{figure}}
\renewcommand{\thesection}{\thepart .\arabic{section}}
\renewcommand\theequation{A\arabic{equation}}
\renewcommand{\thesubsection}{\arabic{subsection}}
\renewcommand{\thesubsubsection}{\alph{subsubsection}}


%


\end{document}